\documentclass[useAMS,usenatbib]{mn2e}
\usepackage{graphicx}
\usepackage{amsmath}
\usepackage{amssymb}
\usepackage{deluxetable}

\def\snx{SN\,2010X} 
\def\Msun{$M_\odot$} 

\newcommand{\kms}{\ensuremath{\mathrm{km~s}^{-1}}}
\newcommand{\Nifs}{\ensuremath{^{56}\mathrm{Ni}}}
\newcommand{\Cofs}{\ensuremath{^{56}\mathrm{Co}}}
\newcommand{\Fefs}{\ensuremath{^{56}\mathrm{Fe}}}
\newcommand{\texp}{\ensuremath{t_{\mathrm{exp}}}}
\newcommand{\tsn}{\ensuremath{t_{\mathrm{sn}}}}
\newcommand{\msun}{\ensuremath{{\rm M}_\odot}}
\newcommand{\Mej}{\ensuremath{M_{\rm ej}}}

\title[Rapidly Fading Supernovae]{Rapidly Fading Supernovae from Massive Star Explosions}

\author[Kleiser et al.]
{Io K. W. Kleiser$^{1}$\thanks{E-mail:io.kleiser@berkeley.edu},
and Daniel Kasen$^{2,3}$\thanks{E-mail:kasen@berkeley.edu},
\\
$^{1}$Department of Astronomy, California Institute of Technology, Pasadena, CA 91125.\\
$^{2}$Lawrence Berkeley National Laboratory, 1 Cyclotron
  Road, Berkeley, CA 94720.\\
$^{3}$Department of Physics, University of California,
  Berkeley, CA 94720.\\}
\begin{document}
\maketitle
\label{firstpage}
\begin{abstract}
Transient surveys have recently discovered a class of supernovae (SNe) with extremely rapidly declining light curves. These events are also often relatively faint, especially compared to Type Ia SNe.  The common explanation for these events involves a weak explosion, producing a radioactive outflow with small ejected mass and kinetic energy ($M \sim 0.1~\msun$ and $E \sim0.1$ B, respectively), perhaps from the detonation of a helium shell on a white dwarf.  We argue, in contrast, that these events may be Type~Ib/c SNe with typical  masses and energies ($M \sim 3~\msun$, $E \sim 1$ B), but which ejected very little radioactive material.   In our picture, the  light curve is powered by the diffusion of thermal energy deposited by the explosion shock wave, and the rapid evolution is due to  recombination, which reduces the opacity and results in an ``oxygen-plateau" light curve.  Using a radiative transfer code and simple 1D ejecta profiles, we generate synthetic spectra and light curves and demonstrate that this model can reasonably  fit the observations of one event, \snx.   Similar models may explain the features of other rapidly evolving SNe such as SN~2002bj and SN~2005ek.  SNe such as these require stripped-envelope progenitors with rather large  radii ($R \sim 20~R_\odot$), which may originate from a mass loss episode occurring just prior to explosion.
\end{abstract}


\section{Introduction \label{intro}}

As more powerful wide field optical surveys come online, not only have the rates of supernova (SN) discoveries increased,  but so has our ability to detect rarer events at greater distances and with lower luminosities.  Of particular interest is a small but growing collection
 of unusual supernovae whose light curves are relatively dim and of short duration. These rapidly fading
supernovae (RFSNe) not only have peculiar light curves, but their spectra are also often distinctive, in some cases containing line features that
have not yet been securely identified. 
Presumably these transients have something interesting to tell us about the life and death of stars, but we still 
do not have a complete  understanding of their physical properties or origins. 

The class of RFSNe is diverse and may be broken up into several subclasses. 
In this paper, we focus on  \snx\  \citep{kasliwal10} and similar events, which have been found in spiral galaxies and so could potentially be related to massive star death. The peak absolute magnitude of \snx\ was $-17$ (corresponding to a luminosity of $\sim  10^{42} ~{\rm erg}~{\rm s}^{-1}$) and the light curve declined very rapidly after peak  (by 0.23 $\pm$ 0.01 mag day$^{-1}$).    The spectra showed  line features of oxygen, calcium, and iron, with some uncertain features attributed to perhaps aluminum or helium.  
The light curve of a seemingly related event, SN~2002bj \citep{poznanski10}, was about two magnitudes brighter than \snx\ but declined at a nearly equal rate.
The spectra of SN~2002bj contained features of silicon, sulfur, and what was tentatively identified as vanadium. 
In both cases, the typical ejecta velocities, measured from the blueshift of the absorption lines, were around $5,000 - 10,000 ~\kms$.   Recently \citet{drout13} 
presented detailed observations of SN~2005ek, which strongly resembles SN~2010X in many ways.

Other RFSNe likely belong to distinct sub-classes.
Events like SN\,2005E have been labeled  ``calclium-rich transients''  \citep{perets10,kawabata10,kasliwal12}, as their late time spectra are dominated by calcium emission. 
These SNe have been found in the outskirts of elliptical galaxies with no signs of star formation and therefore
likely  originate from old stellar populations.   Another class of RFSNe, the
``SN\,2002-cx-like", or ``Iax" events,  show some spectroscopically similarities to SNe Ia, but are distinguished by low peak magnitudes (between about -14 and -19 in $V$-band) and low ejecta velocities \citep{foley13}. 

Several physical models have been proposed to explain the RFSNe.  In almost all cases, the rapid evolution  of the light
curves is explained as a consequence of a low ejected mass ($\sim 0.1~\msun$), resulting in a short photon diffusion time through the ejecta.
The ``.Ia" model, for example, considers the detonation of a thin shell of helium that has accreted onto the surface of a carbon/oxygen (C/O) white dwarf \citep{bildsten07, shen10}.   The model is so named because the kinetic energy ($\sim 0.1$~B, where $1~{\rm B} = 10^{51}~{\rm erg}$) as well as the ejected mass and luminosity are each about a tenth of those of a typical Type~Ia supernova (SN~Ia).  This model can reproduce some basic properties of the \snx\ light curves \citep{kasliwal10}.   However, as we discuss later, the model has difficulty reproducing important features of the observed spectra and the shape of the light curve.  In addition, current ``.Ia" models do not reach
the higher luminosities seen in events like SN~2002bj.

Partial explosions of C/O white dwarfs near the Chandrasekhar mass have also been suggested as an origin
of RFSNe.  \citet{kromer13} simulate a centrally ignited deflagration which burns a portion
of the star but does not release enough nuclear energy to completely unbind it.  Instead, a fraction of the mass ($\sim 0.4~\msun$) is ejected with low kinetic energy and a \Nifs\ content of $\sim 0.1~\msun$. The resulting transients are dim but have fairly long diffusion times due to the relatively high amount of ejected matter and low energy.  The light curves therefore do not decline rapidly enough to match the SN~2010X-like events, although this model may explain the SN\,2002cx-like transients.

Another potentially relevant model is the accretion-induced collapse (AIC) of a white dwarf to a neutron star.  In the AIC simulations of \citet{dessart06}, only a very small amount of radioactive material ($\sim 10^{-4}-10^{-3}$~\msun) is ejected.  The resulting transient should therefore be very dim.  The simulations of \citep{fryer09}, however, find  larger radioactive masses  ($\sim  0.05~\msun$) and predict  brighter SNe.  For rapidly differentially rotating white dwarfs, a centrifugally supported disk may form during collapse and subsequently be blown apart, perhaps synthesizing even more \Nifs\ \citep{metzger09, darbha10, abdik10}.    In these models, the ejecta velocities are fairly large, near the escape velocity of the neutron star ($\sim 0.1-0.3c$).  If the WD is  surrounded by
a relatively dense circumstellar medium, the ejecta may be slowed down and the light curves powered in part by shock heating \citep{fryer09, metzger09}.

While the above white dwarf models may explain a subset of the observational class of RFSNe, in this paper we
argue that they cannot explain all such events.  We consider in particular the \snx-like transients and
highlight two observables that may point to a different origin.  The first is the precipitous decline in the post-maximum light curve 
with no sign, at least within the limits of the observations, of a radioactively powered light curve tail at late times.  This  suggests that the amount of radioisotopes 
ejected is quite small.  
The second is the presence of certain strong spectral features, in particular OI.  
In this paper, we show that the OI lines in \snx\ may require a relatively large  mass of ejected oxygen, which is difficult to accommodate along with the rapid decline in brightness in a ``.Ia" or similar model.    A similar estimate of the oxygen mass is discussed by \citet{drout13} in an analysis of SN\,2005ek.

Some previous studies have considered a core collapse explanation for RFSNe.
\citet{moriya10} simulate low-energy explosions in massive stars and show that, with proper tuning, only a small amount of mass may be ejected ($\sim 0.1~\msun$) with most of the star falling back onto a compact remnant (a black hole).   The 1D models of \citet{moriya10} assume an artificial complete mixing, although the authors speculate that a jet-powered explosion may be able to carry \Nifs\ to the surface layers.
\citet{drout13} also discuss the possibility of a low-energy ( $0.25-0.52 ~{\rm B}$) core collapse explanation for SN~2005ek with an inferred  ejecta mass of $0.3-0.7~\msun$, a \Nifs\ mass of $0.03 \msun$, citing fallback among a few explanations for such a low-mass ejection. One question this raises is how the radioactive \Nifs, which is usually produced in the dense innermost regions of the star, avoids falling back and instead is ejected with the outer layers of the star.

The common feature of all  of the above models has been an unusually low ejected mass and energy.  Here, in contrast, we show that the mass and energy of \snx-like events may be typical of Type~Ib/c SNe ($M \sim 1-5~\msun, E \sim 1$~B).  We attribute the luminosity not to radioactivity but to the thermal energy deposited by the explosion itself.  The rapid  light curve decline, despite the relatively high ejected mass,  can  be explained by recombination,  which dramatically reduces the effective opacity.

The possibility that some SNe may fail to eject radioactive isotopes has
been considered before. \citet{fryer09} discuss models of very massive stars ($\gtrsim 20~\msun$) in which the amount of material that falls back onto the remnant may be quite substantial, i.e. several solar masses.  Essentially all of the $^{56}$Ni is formed in the innermost layers of the ejecta and falls back, robbing the light curves of energy from radioactive decay and producing very dim events ($V$ and $B$ magnitudes of $-13$ to $-15$).   \citet{ugliano12} indicate that such a large amount of fallback material is unlikely, probably not more than $\sim 0.2 \msun$, but this may still be enough to accrete most if not all of the radioactive material produced. \citet{dessart11} have  considered Type~Ib models that lack \Nifs\ and show that they produce relatively short duration, thermally powered light curves with  peak luminosities ($\sim 10^{40} - 10^{41}~{\rm ergs}~{\rm s}^{-1}$).

In this paper, we explore such models of massive star explosions as an explanation  for  SN~2010X-like events.
We first argue that SN~2010X ejected a substantial amount of oxygen, suggestive  of the explosion of a stripped-envelope, massive star (\S\ref{o_constraint}).  We then demonstrate that massive SNe can produce brief, rapidly declining light curves once recombination is taken into account (\S\ref{sec:opacity}).  We then 
use the radiative transfer code SEDONA \citep{kasen06} to produce synthetic light curves and spectra of simple, 1-D ejecta models, and show that
these models can reproduce the \snx\ light curve, provided that
the progenitor star had a large enough radius (\S\ref{sec:sedona}).  In  \S\ref{discussion}, we discuss
various progenitor scenarios that may be responsible for this class of SNe.

\section{Estimates of the Ejecta Mass}
 \label{o_constraint}

Analytical scaling relations are commonly used to estimate the ejecta mass and kinetic energy of observed SNe.   For ejecta of mass $M_{\rm ej}$ and velocity $v$, and assuming a constant opacity $\kappa$, the duration of the light curve \tsn\ is set by the effective diffusion time through the expanding ejecta \citep{arnett79}:
\begin{equation}
t_{\rm sn} \approx 34
\biggl( \frac{M_{\rm ej}} {\msun}  \biggr)^{1/2}
 \kappa_{0.1}^{1/2}  v_4 ^{-1/2}~{\rm days},
 \label{eq:diff_time}
\end{equation}
where  $v_4 = v/10^4~\kms$ and  $\kappa_{0.1} = \kappa/0.1~{\rm cm^2~g^{-1}}$.  We have calibrated the numerical constant based on Type~Ia SNe,  which have $v_4 \approx 1, M_{\rm ej} \approx 1.4~\msun$ and a bolometric light curve width (i.e., rise plus fall) of  roughly 40 days \citep{contardo00}.   The value $\kappa = 0.1~{\rm cm^2~g^{-1}}$ is  appropriate for electron scattering  in singly ionized helium.  It is also similar to the mean opacity due to Doppler broadened lines of iron-group elements, which is the dominant form of opacity in SNe~Ia \citep{pinto00}.  

Inverting Equation \ref{eq:diff_time} for the ejecta mass gives
\begin{equation}
M_{\rm ej} \approx 0.0875~\biggl(\frac{t_{\rm sn}}{10~{\rm days}} \biggr)^2  v_4 \kappa_{0.1}^{-1}
~\msun,
\label{eq:inferred_mass}
\end{equation}
which has fostered the belief that RFSNe like  \snx\  represent relatively low mass ejections.  Such an argument, however, presumes a constant opacity of $\kappa \sim 0.1~{\rm cm^2~g^{-1}}$.   In fact, the opacity of SN ejecta is highly dependent on the
physical state and, as we discuss in the next section, may vary by an order of magnitude depending on the temperature and composition of the ejecta.

It is possible to derive an independent constraint on the ejecta mass using absorption features observed in the  spectrum.  In particular, \snx~showed a strong, broad, and persistent OI triplet feature ($\lambda\lambda\lambda$7772,7774,7775), which would seem to suggest a significant mass of oxygen.   
In homologously expanding atmospheres, the degree of absorption is quantified by the Sobolev optical depth,\begin{equation}
\tau_{\rm sob} = \biggr( \frac{\pi e^2}{m_e c^2} \biggl) \texp \lambda_0 n_l,
\end{equation}
where $\lambda_0$ is the rest wavelength of the line, 
\texp\ the time since explosion, and  $n_l$ the number density in the lower level of the atomic transition.  

\begin{figure*}
\begin{center}
\includegraphics[width=7in]{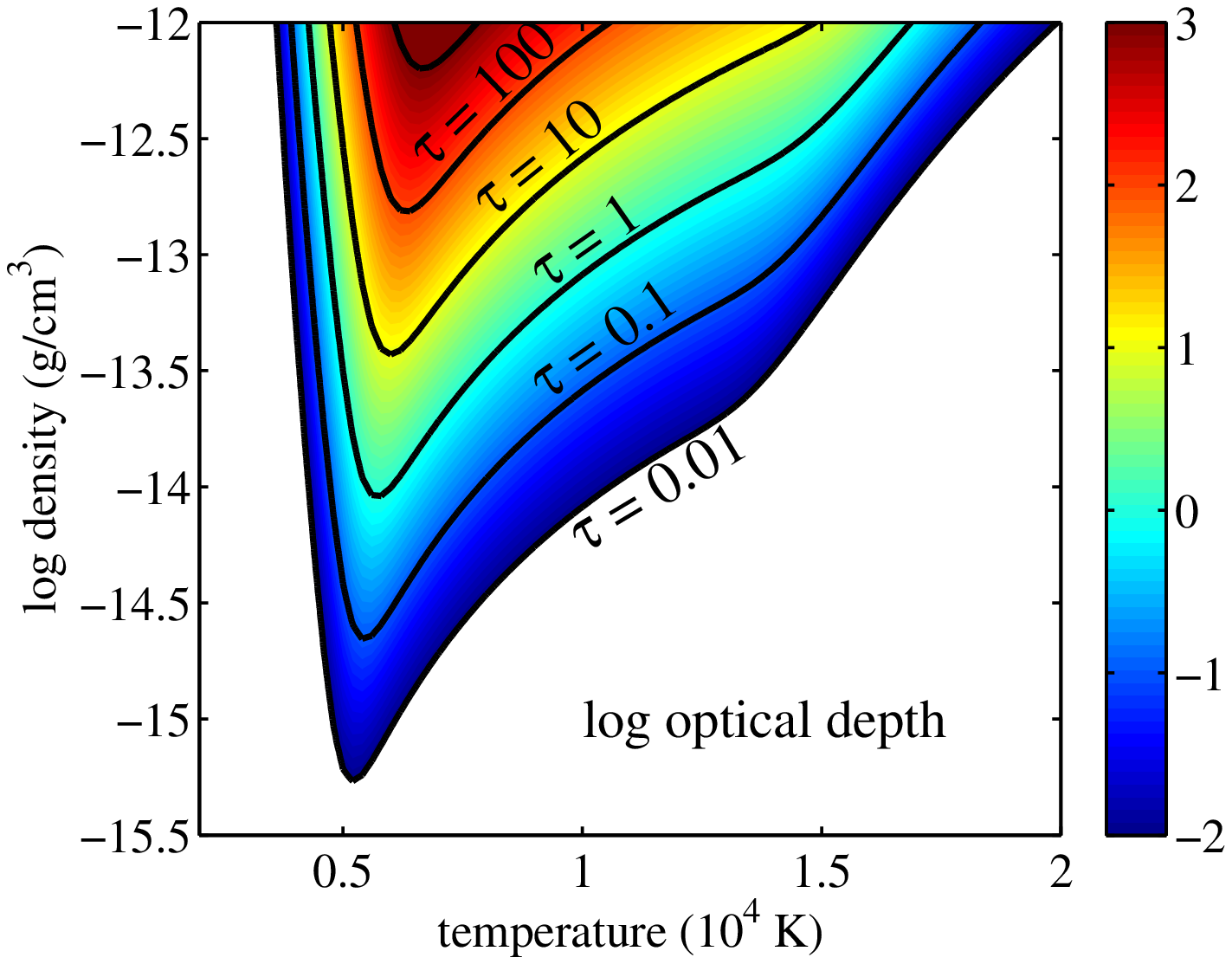}
\end{center}
\caption{Logarithmic Sobolev optical depth of the combined OI  $\lambda\lambda\lambda$7772,7774,7775 triplet line as a function of  density and temperature for a 100\% oxygen composition at \texp = 30~days. Black lines indicate curves of constant optical depth. The lowest point in the $\tau = 1$ curve occurs at $\rho_c \approx 10^{-14}$~g/cm$^3$, which is taken to give the lowest possible density needed to see the OI absorption. \label{tau}}
\end{figure*}

In Figure \ref{tau}  we plot  contours of $\tau_{\rm sob}$ for the OI triplet line as a function of  density and temperature for ejecta composed of pure oxygen at  $\texp = 30$~days, a time when the  oxygen absorption feature is quite prominent in SN~2010X.
To estimate $n_l$ we have assumed that the ionization/excitation states were approximately  given by local thermodynamic equilibrium (LTE).  
Figure \ref{tau} shows that a density of at least  $\rho_{\rm c} \approx 10^{-14}~{\rm g~cm^{-3}}$ is required to achieve  strong ($\tau_{\rm sob} \ga 1$) absorption in the OI line.  This  critical density  corresponds to the most favorable temperature, $T \approx 5500$~K, at which the level density $n_l$ is highest.   For hotter temperatures, oxygen becomes more highly ionized,  while for cooler temperatures it is difficult to thermally populate the excited lower level of the OI transition.  In these cases, an 
even higher density is required to make the  line optically thick.

We  can use this critical OI density to obtain an approximate lower limit on the total ejecta mass of SN~2010X. The February~23 spectrum (close to $t_{\rm exp} \approx 20$~days) showed  apparent OI absorption at velocities $\approx 10,000~\kms$.   Assuming that the ejecta density profile is described  by a broken power-law  (see \S\ref{sec:sedona})  and that the energy release per unit mass is typical of SNe,  $E/M \sim 10^{51}~{\rm ergs}/\msun$, the condition  $\rho \gtrsim \rho_c$ at $v \approx 10,000~\kms$ implies a total ejecta mass   $M_{\rm ej} \gtrsim 0.35~\msun$.   This is likely a significant underestimate,  as we have assumed the oxygen layer was  composed of 100\% oxygen at the ideal temperature of $T \approx 5500$~K. The March 7 spectrum of SN~2010X (near $t_{\rm exp} \approx 30$~days) also shows strong absorption at the same location, which implies $M_{\rm ej} \gtrsim 1.2~\msun$.

These spectroscopic mass estimates are subject to two important caveats.  First, the absorption near $7500$~\AA\ may be due 
in part to MgII (and perhaps FeII) rather than OI lines.  Second, the line optical depths may be influenced by non-LTE effects.  The lower level of the OI feature is a quasi-stationary state (albeit with a high excitation energy, $\Delta E = 9.14$~eV) 
such that an LTE description may not be a bad approximation.  Nonetheless, it is quite possible
that the level population is enhanced by non-thermal excitation by radioactive
decay products \citep{Lucy_1991,Dessart_2012}, or by time-dependent effects \citep{Dessart_2008}.
As an empirical  check of the method, one can examine the constraints  for normal SNe~Ia.  For SN~1994D, for example, the OI feature remained mildly optically thick ($\tau_{\rm sob} \sim 1$) until about 12 days past maximum light ($\texp \approx 30$~days).   From this, one calculates a total ejecta mass of $\sim 1-2~\msun$, which is consistent with the values expected for SNe~Ia.  The OI feature of \snx\ at a comparable epoch ($\texp \approx 30$~days)  is significantly broader and deeper than that of SN~1994D.   We therefore consider it likely  that the ejected mass of \snx\ was comparable to or larger than that of a typical SNe~Ia,
hence $M_{\rm ej} \gtrsim 1~\msun$. 

\begin{figure*}
\begin{center}
\includegraphics[width=7in]{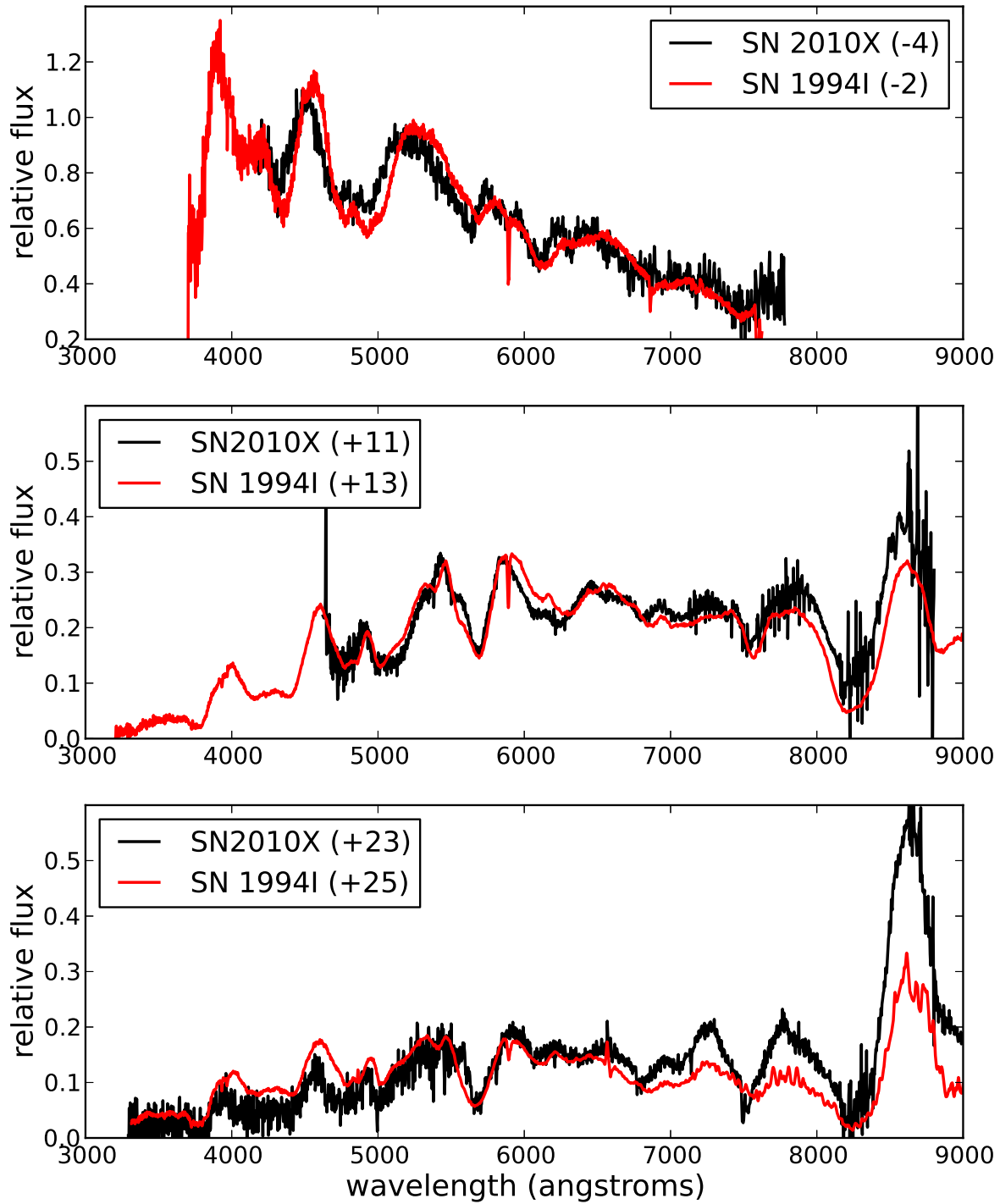}
\end{center}
\caption{Comparison of the multi-epoch spectra of the Type Ic  SN~1994I to those of SN~2010X.  Times since B-band maximum are listed.  The strong spectral similarity may indicate a similar physical origin.  \label{94i}}
\end{figure*}

The association of SN~2010X with a massive progenitor is strengthened by comparison with  core-collapse SNe. Figure~\ref{94i} shows the SN~2010X spectra with those of SN\,1994I, which is considered a fairly typical, if somewhat fast-evolving, Type~Ic SNe.  The agreement is striking and strongly points to a similar physical origin for the two.  \citet{drout13} have also noted the spectroscopic resemblance of SN~2005ek to other normal SNe Ic as well as \snx.    The light curve of SN~1994I showed a clear radioactive tail, indicating that it ejected $\sim 0.07~\msun$ of \Nifs\ \citep{young95,iwamoto94}.  We will suggest that \snx~was a compositionally similar Type Ib/c SN but did not eject as much \Nifs.

\section{Oxygen Plateau Supernovae}
\label{sec:opacity}

The mass estimates discussed in the last section present a paradox -- the narrow light curve of \snx\ suggests a low \Mej, while  the spectroscopic constraints indicate that \Mej\ may be many times larger.
Here we show that the conflicting estimates can be reconciled in a core-collapse model
in which the ejecta mass is large ($M \gtrsim 1~\msun$) but where the effective diffusion
time is significantly reduced due to recombination.

\begin{figure*}
\begin{center}
\includegraphics[width=7in]{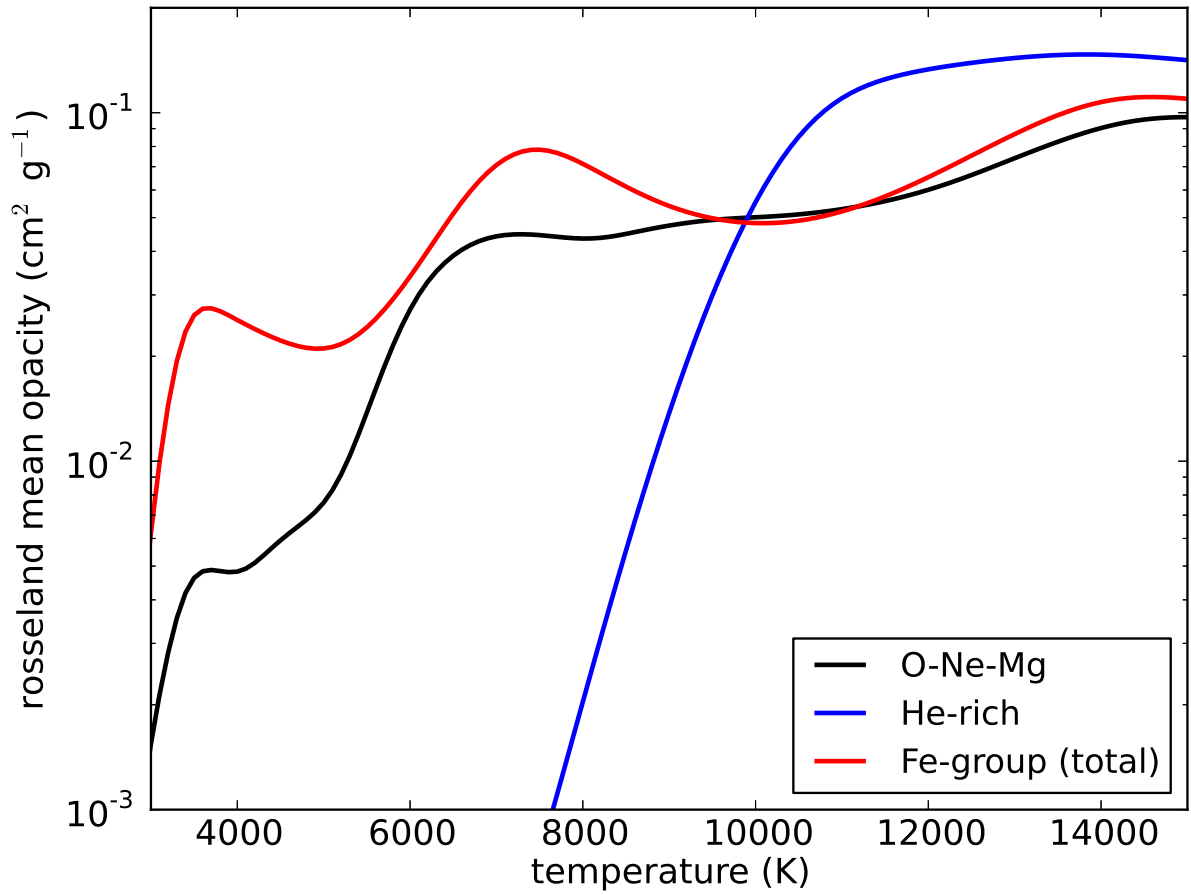}
\end{center}
\caption{Calculated Rosseland mean opacity (for SN ejecta of different compositions) as a function of temperature for supernova ejecta at a density $\rho = 10^{-13}~{\rm g~cm^{-3}}$ and $\texp = 10$~days.   The main opacities included are electron scattering and line expansion opacity. \label{fig:opacity}}.
\end{figure*}

The opacity of SN ejecta is highly dependent on the ionization state and so may vary significantly  with temperature.
In Figure~\ref{fig:opacity}, we plot the Rosseland mean opacity (calculated assuming LTE) of SN ejecta of different compositions.  We consider in particular a oxygen-neon-magnesium composition (see Table~1) which may be characteristic of the massive, stripped envelope stars believed to be the progenitors of Type~Ic SNe.    For higher temperatures ($T \ga 6000$~K), the O-Ne-Mg opacity  has a characteristic value  $\kappa \approx 0.04~{\rm cm^2~g^{-1}}$  When the temperature drops below $6000$~K,  however, oxygen recombines to neutral, and the opacity drops sharply by more than an order of magnitude.    This is because, in the absence of scattering off of free-electrons, photons can escape through  the ``windows" in wavelength space that occur between the lines, reducing the Rosseland mean opacity. However, the opacity does not drop to zero at $T \lesssim 6000$~K because other elements (such as Mg and Si) with lower ionization potentials remain ionized.  For a helium-rich composition, recombination occurs at a higher temperature ($\sim 10,000~K$) due to the higher ionization potential of helium. In contrast, the opacity of \Nifs\ and its daughter nuclei (\Cofs\ and \Fefs), due to the lower ionization potential of the iron group species, maintains a large value as long as $T \ga 3000$~K.

The recombination physics will strongly influence the light curves of SNe  
composed largely of oxygen. 
The radiative transfer parallels the well understood effects in Type~II plateau SNe \citep[e.g.,][]{Grassberg_1971, Dessart_2008}.  Initially, the ejecta are heated and ionized by the passage of the explosion shockwave.  As the ejecta expand and cool, however, the 
material eventually drops below the recombination temperature $T_{\rm i}$ and becomes largely transparent. Because the outer layers of ejecta are the coolest, they recombine first, and a sharp ionization front  develops in the ejecta.   As time goes on, the ionization front  recedes inward in mass coordinates, releasing the stored  thermal energy.   When this recombination wave reaches the center of the ejecta, the stored energy is exhausted and the light curve should drop off 
very rapidly, marking the end of the ``oxygen-plateau" phase.   The analogous case of a helium plateau in Type~Ib SN has
been discussed by \citet{ensman88} and \citet{dessart11}.

Relationships for the timescale and peak luminosity of Type~II plateau supernovae, including the effects of recombination, have been determined by \citet{popov93} and verified numerically by \citet{kasen09}, who find
\begin{equation}
t_\mathrm{sn} \approx 120 E_{51}^{-1/6}M_{10}^{1/2}R_{500}^{1/6}\kappa_{0.4}^{1/6}T_{6000}^{-2/3}~{\rm days}\,\,,
\label{eq:t_sn}
\end{equation}
\begin{equation}
L_\mathrm{sn} \approx 1.2 \times 10^{42} E_{51}^{5/6}M_{10}^{-1/2}R_{500}^{2/3} \kappa_{0.4}^{-1/3}T_{6000}^{4/3}~{\rm ergs/s}\,\,,
\label{eq:L_sn}
\end{equation}
where $E_{51} = E / 10^{51}~\rm{ergs}$ is the explosion energy, $M_{10} = M /(10$ \Msun) is the ejected mass, $R_{500} = R /(500~R_\odot)$ is the presupernova radius, $\kappa_{0.4} = \kappa /(0.4~\rm{cm}^{2} \rm{g}^{-1})$ is the opacity of the ejecta, and $T_{6000} = T /(6000~\rm{K})$ is the ejecta temperature. The numerical calculations of \citet{kasen09} actually found a scaling closer to $t_{\rm sn} \propto E^{-1/4}$ rather than $E^{-1/6}$, but otherwise the relations are the same.
 We can use similar arguments for hydrogen-less supernovae, assuming the luminosity is determined by the recombination temperature of whatever species dominates the ejecta.

Inverting Equations \ref{eq:t_sn} and \ref{eq:L_sn} allows us to solve for the ejecta mass and presupernova radius in terms of observed quantities,
\begin{equation}
M_{\rm ej} \approx 2.9~ L_{42}^{-1} t_{20}^4  v_4^3 \kappa_{0.04}^{-1} T_{6000}^4~ \msun\,\,,
\label{eq:m_estimate}
\end{equation}
\begin{equation}
R_0 \approx 12.4~ L_{42}^{2} t_{20}^{-2}  v_4^{-4} \kappa_{0.04} T_{6000}^{-4}~ {\rm R}_\odot\,\,.
\label{eq:r_estimate}
\end{equation}
These are very rough estimates, but they demonstrate that, when recombination is accounted for,
the light curves of SN~2010X and other RFSNe are consistent with massive ($M_{\rm ej} \gtrsim 1~\msun$) ejections which  powered
by shock energy, not radioactivity, provided that the radius of the progenitor star
is sufficiently large.

\section{Radiative Transfer Models}
\label{sec:sedona}

To model the light curves and spectra of RFSNe in more detail, we use the time-dependent 
Monte Carlo code radiative transfer code SEDONA \citep{kasen06}.  
We base our calculations on simple parameterized  ejecta models rather than
on detailed hydrodynamical simulations, as this  allows us to easily control the ejecta mass,
kinetic energy, and progenitor star radius.  We vary these parameters, in an empirical spirit,
in an effort to fit the observations of SN~2010X and  constrain its physical properties.

\subsection{Ejecta Models}

For simplicity, we consider ejecta models that are spherically symmetric and in the homologous expansion phase.
Simulations of core-collapse explosions suggest that the ejecta density structure can roughly
be described by broken power-law profile \citep{chevalier92},
\begin{equation}
\begin{split}
\rho_\mathrm{in}(r,t) = \zeta_\rho \frac{M}{v_\mathrm{t}^3 t^3} \frac{r}{v_\mathrm{t} t}^{-\delta}~{\rm for}~v < v_t\,\,,
\\
\rho_\mathrm{out}(r,t) = \zeta_\rho \frac{M}{v_\mathrm{t}^3 t^3} \frac{r}{v_\mathrm{t} t}^{-n}~{\rm for}~v \geq v_t\,\,,
\end{split}
\label{eq:dens_profile}
\end{equation}
where $v_\mathrm{t}$ is the velocity at the transition between the two regions, 
\begin{equation}
v_\mathrm{t} = 4.5 \times 10^8 \zeta_v(E_{51}/\msun)^{1/2} \mathrm{cm}~\mathrm{s}^{-1}\,\,.
\end{equation}
The coefficients $\zeta_\rho$ and $\zeta_v$ are constants which can be determined by requiring Equation~\ref{eq:dens_profile} integrate to the specified mass and energy. In our model for \snx, we use $\delta=1$ and $n=8$ as they are typical values for core collapse SN and produced reasonable fits to the light curves and spectra.

Immediately following the passage of a core-collapse SN shockwave,
 the explosion energy is roughly equally split between the kinetic energy and thermal energy of the stellar material.  The latter is strongly radiation
dominated.
Simulations suggest that, before radiative diffusion sets in, the ratio of the radiation energy density 
to the mass density is nearly constant throughout most of the envelope \citep{woosley88}.
 We therefore take the energy density profile at $t_0$,  the start time of our calculation, to be
\begin{equation}
\epsilon(v,t_0)  = \frac{E_0}{2} \frac{\rho(v)  }{ M} \biggl( \frac{R_0}{R_{\rm ej}} \biggr)\,\,,
\end{equation}
where $R_{\rm ej} = v_t t_0$ is the size of the  remnant at the start of our transport calculation.  
This expression assures that the total thermal energy equals $E_0/2$ when $R_{\rm ej} = R_0$; 
the term in parentheses accounts for losses due to adiabatic expansion prior to the start of
 our transport calculation.  In this model, the initial energy density profile is
 a broken power law with the same exponents as the mass density.   
This is reasonably consistent with analytical results that find that the energy density power law in the outer layers is very similar to, though slightly steeper than, the mass density profile \citep{chevalier92}.

We assume that the composition of the ejecta is homogenous in two layers. 
Detailed abundances are given in Table \ref{t:comp}. Abundances for the inner layers ($v \le 10000$~\kms) are typical of an O-Ne-Mg layers of a massive star  and  taken from  the stellar evolution models
of  \citet{woosley02} for a 25~\msun\ pre-supernova star at a mass coordinate of $3.9~\msun$.   
For the outer layers ($v > 10000$~\kms) we assume He-rich material with a solar abundances of metals.  The inclusion of a helium in the outer layers in fact does not significantly affect the light curves and spectra, as the photosphere at the epochs of
interest turns out to be in the O-Ne-Mg layers.  
For the models in this paper, we assume that no radioactive isotopes were ejected, so  the light curves are solely powered by the energy deposited in the explosion shock wave.

\subsection{Synthetic Light Curves and Spectra}

\begin{figure*}
\begin{center}
\includegraphics[width=7in]{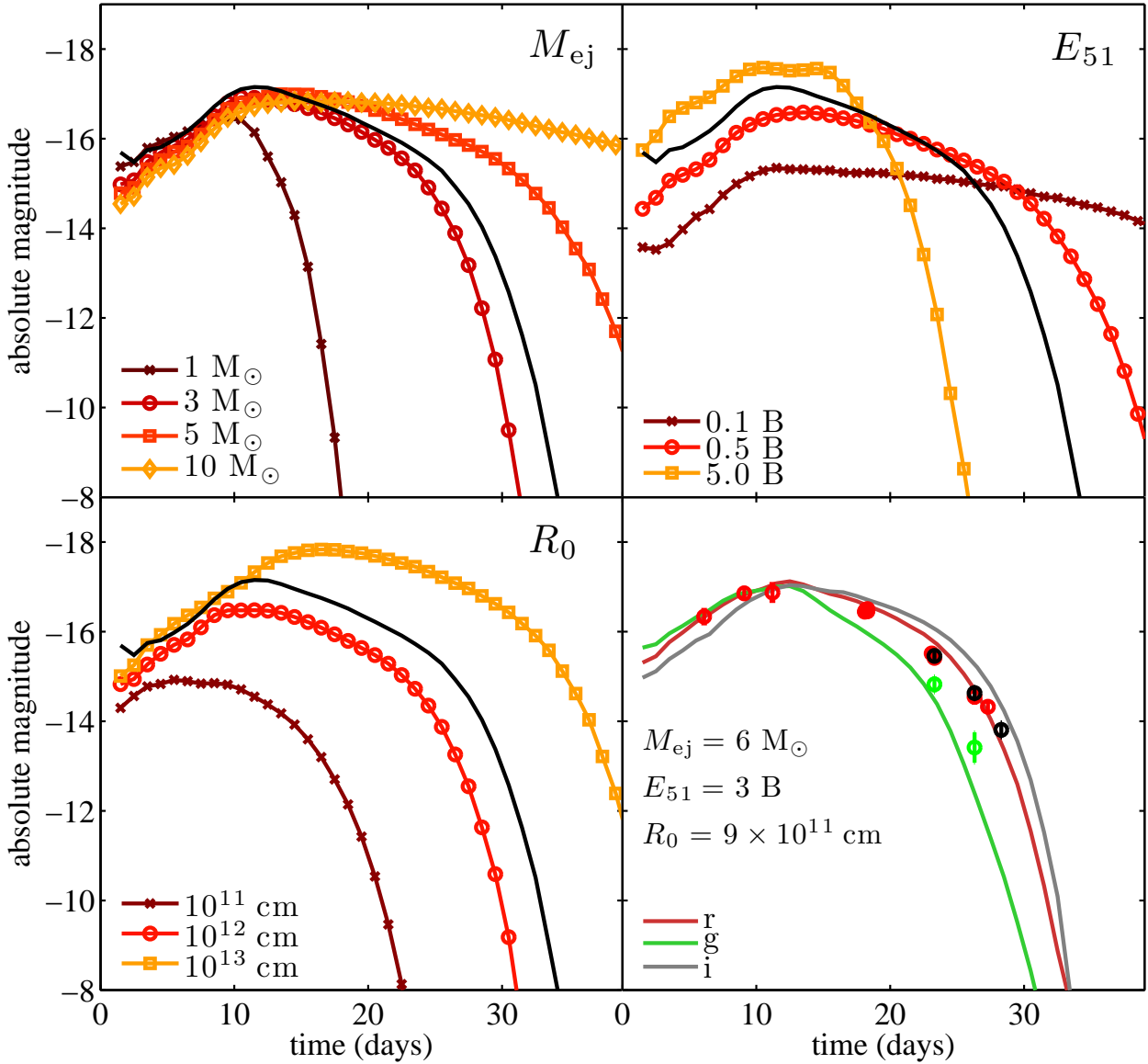}
\end{center}
\caption{ Calculated light curves using parameter variations around our fiducial ejecta model for \snx, which has parameters $M_{\rm ej} = 3.5~\msun$, $E_{51} = 1~{\rm B}$, and $R_0 = 2 \times 10^{12}~{\rm cm}$. Top left: light curve calculations holding all parameters constant except ejecta mass. Top right: same as the top left panel but with varying explosion energy. Bottom left: same as top right and top left panels but with varying presupernova radius. Bottom right: an alternative model that fits the data fairly well with parameters $M_{\rm ej} = 6~\msun$, $E_{51} = 3~{\rm B}$, and $R_0 = 9 \times 10^{11}~{\rm cm}$.  This demonstrates the degeneracy in our approach and that the light curves could be fit with a range of parameters. \label{varyparam}}
\end{figure*}

We  performed radiative transfer calculations for a series of models, in which we vary the three key ejecta
parameters:  the explosion $E$, the  ejected mass $M_\mathrm{ej}$, and the presupernova radius $R_0$.
Figure \ref{varyparam} shows how the SDSS r-band light curve changes as we vary each parameter  while holding the others fixed. We show the r-band curves for easy comparison to SN\,2010X, as this is the band in which we have the most data.
The general trends are qualitatively consistent with the scaling relations (Equations \ref{eq:t_sn} and \ref{eq:L_sn}).
Increasing the explosion energy  shortens the light curve duration while increasing the peak luminosity. Raising the mass increases the light curve duration but does not strongly affect its peak luminosity. Finally, a larger presupernova radius increases both the luminosity and duration of the light curve. 
 The light curves resemble those presented in \citet{dessart11} for models 
of Type~Ib/c SNe assumed to eject no \Nifs.

\begin{figure*}
\begin{center}
\includegraphics[width=7in]{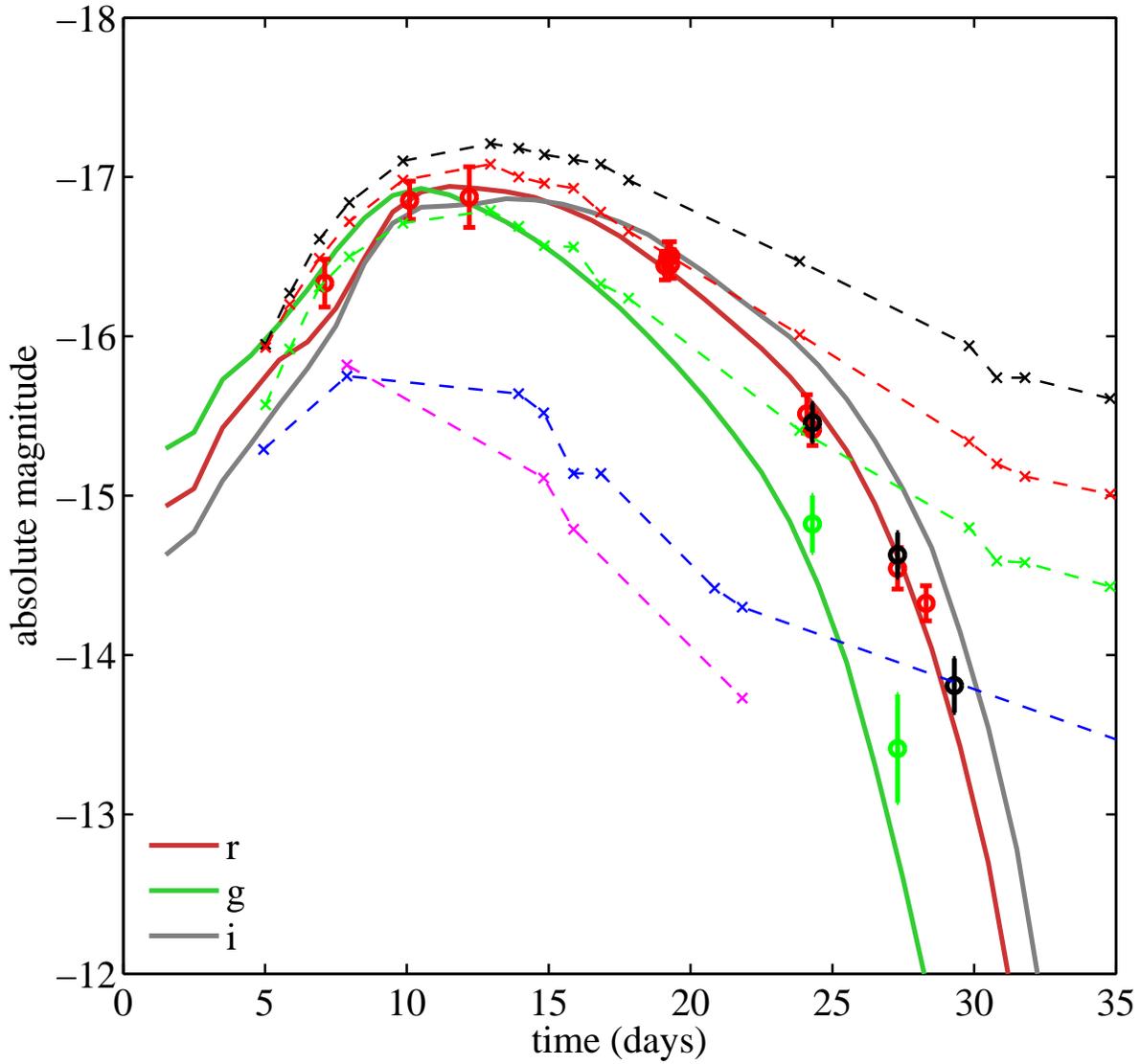}
\end{center}
\caption{Light curves in $g$, $r$, and $i$ calculated for a pure explosion model of \snx, plotted against the data. This model was obtained with $M_{\rm ej} = 3.5~\msun$, $E_{51} = 1~{\rm B}$, and $R_0 = 2 \times 10^{12}~{\rm cm}$. Dotted lines show light curves in $BVRI$ for SN\,1994I, a typical SN\,Ic, for comparison. \label{snx_lc}}
\end{figure*}

In Figure \ref{snx_lc} we show a fit to the light curve of SN~2010X, 
using a model with $M_{\rm ej} = 3.5~\msun, E = 1~{\rm B}$, and $R_0 = 2\times 10^{12}$ cm. 
The model demonstrates that the basic properties of this RFSNe can be explained by the explosion of an ordinary-mass star in which the emission is powered solely by the energy deposited in the explosion shockwave, without any radioactive \Nifs.  The assumed radius of the progenitor, however, is significantly larger than that of typical Wolf-Rayet stars, an issue we return to in \S\ref{discussion}.
The short duration of the model light curve reflects the
rapid release of radiation energy by the receding recombination wave.  The luminosity for the first 25~days (the ``oxygen-plateau") is fairly constant,
but then drops dramatically as  the recombination wave nears the center of the ejecta and the stored radiation energy is  exhausted.  After day 25, the r-band magnitude drops by more than 3 magnitudes in only 5 days,
marking the end of the plateau phase.  As no radioisotopes were included, the light curve shows no radioactive tail at late times, and the luminosity continues to drop rapidly.

The light curve fit does not uniquely constrain all three model parameters ($M, E$, and $R_0$ ) as there are essentially only two photometric observables (light curve brightness and duration).  We have chosen to show here a model in which
the mass and energy are typical of ordinary Type~Ic SNe, but other combinations can provide fits of similar quality (see Figure~\ref{varyparam}).   The degeneracy can perhaps be broken by using the observed velocity to constrain the mass energy ratio; however, the photospheric velocity in plateau SNe is set by the location of the recombination front and hence is not necessarily indicative of  $v \approx (2 E/M)^{1/2}$.

\begin{figure*}
\begin{center}
\includegraphics[width=7in]{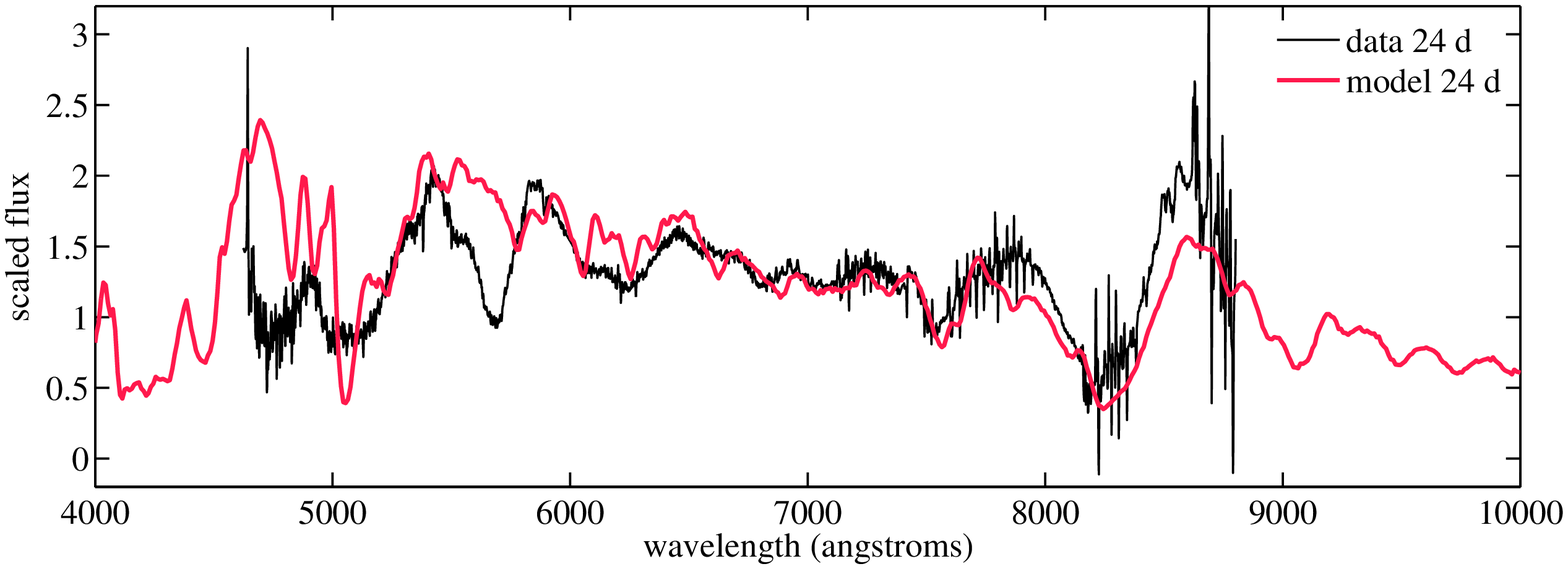}
\end{center}
\caption{Selected spectrum calculated from our fiducial ejecta model of Figure~\ref{snx_lc} shown against observed data. The overall shape is similar, and most of the important spectral features are reproduced. Discrepancies may arise from our assumption of LTE,  simplified power-law density structure, or the untuned abundances assumed. \label{snx_bestspec}}
\end{figure*}

Figure~\ref{snx_bestspec} compares the synthetic spectrum (at $\texp = 24$~days) of the same model to the February~23 spectrum of SN~2010X.  On the whole, the 
model does a good job reproducing the major spectral features 
and in particular predicts significant absorption near the OI triplet. 
This supports the idea that the composition of the SN~2010X ejecta is consistent with that of a O-Ne-Mg core
of a massive star.   In detail, however, one notices discrepancies in the position and depth of several features. For instance, the model absorption near 5600~\AA, due to the sodium NaID line, is much too
weak and has too low a velocity.  A similar problem with the NaID line has often been noted in models of Type~IIP SNe, and has been explained as
resulting from the neglect of time-dependent non-LTE effects \citep{Dessart_2008}.
Thus, while fine-tuning of our ejecta parameters could likely improve the spectral
fit,  the overall agreement is presumably limited by the simplified nature of the calculations,  including the one-dimensional broken power law density structure, 
the two-zone uniform composition, and the neglect of non-LTE effects.

The identification of the absorption features at $6800$~\AA\ and $7000$~\AA\ was the subject of some discussion in \citet{kasliwal10}, who suggest that these features may be due either to lines of AlII or HeI.  
Our model does not include aluminum, and the helium lines are optically thin, given the lack of non-thermal excitation from radioactivity.  Analysis of the Sobolev  optical depths suggest that lines of FeII and neutral species (SiI~$\lambda7035$ and CaI) contribute to the spectral features in this wavelength
region. \citet{drout13}  similarly show that the spectra of SN~2010X-like 
events can be reasonably fit without invoking aluminum or helium absorption lines.

\begin{figure*}
\begin{center}
\includegraphics[width=7in]{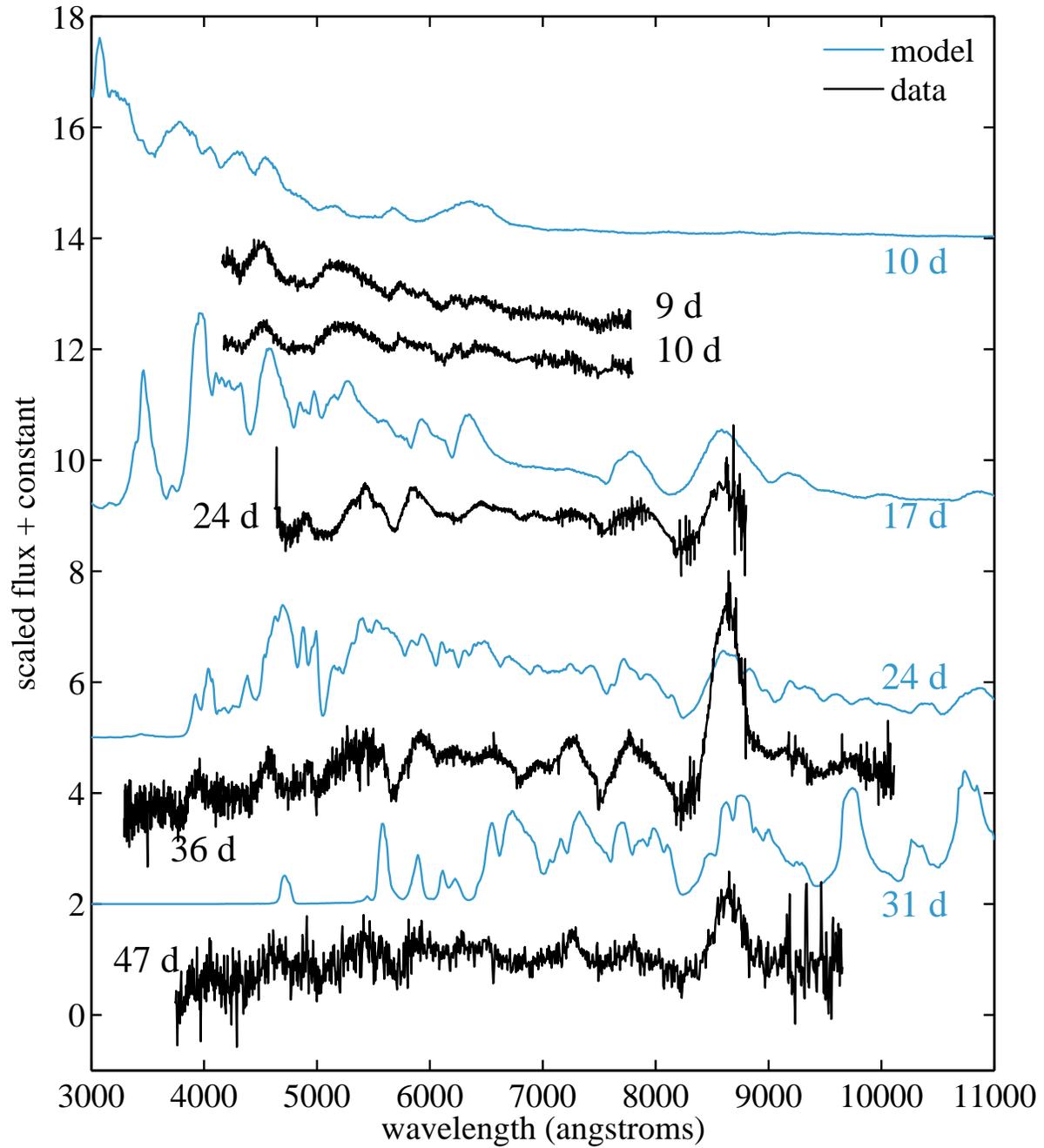}
\end{center}
\caption{Time series of selected synthetic spectra of our fiducial ejecta model of Figure~\ref{snx_lc} compared the observed data of \snx~showing the evolution of the oxygen line and other prominent features. \label{snx_timeseries}}
\end{figure*}

In Figure~\ref{snx_timeseries}, we show the spectral time series of  SN~2010X  alongside select spectra from our model.   The general trends are reasonable, 
but the color evolution is faster in the model.  For example, the day~10 model spectrum is bluer  than the day~9 observed spectrum, while the day~31 model spectrum is redder than the observed day~36 spectrum.  
Our radiative transport becomes suspect at later times ($\gtrsim 30$~days) as the ejecta are becoming optically thin and non-LTE effects should become more significant.  
While the model spectral series does not reproduce every observed spectral feature, we emphasize that we have chosen to limit any fine-tuning of the abundances and explosion parameters in order to fit the data.  Further adjustment of the oxygen-rich composition would presumably lead to an improved fit, as has been shown in the modeling of the
spectroscopically similar SN~1994I \citep{sauer06}.

\section{Discussion and Conclusions}
\label{discussion}

We have argued that some RFSNe, in particular the SN~2010X-like events, are  the result of core-collapse
explosion of massive stripped-envelope stars.   This contradicts previous suggestions that these events represent low mass, low-energy outbursts from, for example, ``.Ia" explosions on white dwarfs. In our picture, the supernova ejected very little radioactive material 
and the  light curve was instead powered by the diffusion of thermal energy deposited by the explosion shock wave.  The short duration of the light curve, despite the relatively high ejected mass ($M \sim 3.5~\msun$), 
is due to recombination,  which dramatically reduces the effective opacity.    
The evolution is similar to Type~IIP supernovae,  and the sharp decline of the light curve can be understood as reflecting the end of an 
``oxygen plateau".  Our  1D radiation transport models demonstrate that the observations of SN~2010X are consistent with this scenario.  Empirically, the  spectral similarity of SN~2010X with the Type~Ic SN~1994I strongly suggests that these events
have oxygen-dominated ejecta as would be expected in stripped core-collapse SNe.

Other RFSNe may have a similar origin.  The light curve of SN~2002bj had a very similar decline rate to that of SN~2010X, but the peak luminosity was about two magnitudes brighter.  The model scalings suggest that the brightness and duration could be reproduced in a Type~Ib/c plateau SN with larger progenitor radius and/or higher explosion energy.  The spectral features of SN~2002bj were distinct from \snx, perhaps because the ejecta temperatures were higher, but possibly because the composition of the ejecta was
 different (e.g., helium-rich instead of oxygen-rich). Further  modeling is needed to constrain the ejecta properties in detail.

Recently, \citet{drout13}  presented a detailed analysis of SN~2005ek, which was  spectroscopically and photometrically similar to SN~2010X.   
The late-time R and I band observations of SN~2005ek (at 40 and 70 days after peak) do indicate the presence of a radioactively powered
light curve tail.     The uncertain bolometric corrections, however, make it difficult to determine the actual gamma-ray trapping rate and hence radioactive mass.   
The late-time light curve decline is consistent with nearly complete gamma-ray trapping, as might be expected in a massive ($M \sim 3 \msun$) event.  In this case,  the inferred \Nifs\ mass, while not zero, is very small,  
$M_{\rm ni} \approx 1 - 4 \times 10^{-3}~\msun$.  The luminosity at the light 
curve peak would then be attributed to an oxygen-plateau of the sort we
have described.

In contrast, \citet{drout13} argue that if the ejecta mass and kinetic energy of SN~2005ek were relatively low ($M \approx 0.35 - 0.7~\msun, E \approx 2.5-5.2~B$) then most of the gamma-rays escape at late times and the inferred \Nifs\ mass is much larger, $M_{\rm ni} \approx 0.03~\msun$, sufficient to power the  light curve peak. It is not clear, however, that such a model can explain the rapid light curve decline after peak.  \citet{Fink_2013} have calculated radiative transport models for similar scenarios (e.g., model ND3 with $M \approx 0.2~\msun, E \approx 0.43~B, M_{\rm ni} \approx 0.07~\msun$) and find light curves that decline fairly gradually, dropping by only $\sim 1$~mag in R-band in the 15 days after peak.  This is much more gradual than either SN~2005ek or SN~2010X (which dropped  $\sim 3$~R-band mags in 15 days), suggesting that the radioactively powered model may struggle to explain the rapid decline that characterizes this class of SNe.

The light curve of SN~2002bj poses an even greater challenge for radioactively powered models.  This event was significantly brighter than either SN~2005ek and SN~2010X, such that the inferred  \Nifs\ mass would be $M_{\rm ni} \approx 0.15-0.25$~\msun\ \citep{poznanski10}.  To be consistent with the rapid rise and steep decline of the light curve, one would need to assume a small ejecta mass, $M \sim 0.3~\msun$, such that the ejecta consisted of very little else but \Nifs.  This, however, contradicts the observed spectrum, which did not show strong features from iron group elements.

A potentially revealing empirical discriminant of the RFSNe 
is the ratio of the luminosity measured at peak to that on the radioactive tail.   Some events,  like SN~2002cx, SN~2005E and SN~2008ha, show a moderate peak-to-tail ratio, similar to SNe~Ia and consistent with a light curve powered entirely by radioactivity.  In contrast, the events considered here 
(SN~2005ek, SN\,2010X, and SN\,2002bj) show a larger peak-to-tail
 luminosity ratio (or no tail at all) more reminiscent of Type~IIP supernova.
It is possible that this distinction separates those events
powered solely by radioactivity from those with an initial
thermally powered oxygen (or helium) plateau.

The RFSNe also are distinguished by their host galaxies.
All of  SN~2010X,  SN~2002bj, and SN\,2005ek were found in star-forming galaxies and so are consistent with 
young stellar populations and massive star progenitors.  Other types of RFSNe, however, such as SN~2005E and similar low-luminosity calcium-rich transients have often been found in the remote outskirts of elliptical galaxies, which almost exclusively harbor old stars.  For these events, a different model, perhaps based on white dwarf progenitors, may be appropriate.   

If correct, the identification of SN~2010X-like events as oxygen plateau  SNe has two important implications for core-collapse SNe.  The first is that some stripped-envelope SNe may eject a very small amount $(\lesssim 10^{-3}~\msun$) of radioactive isotopes.  This may be because abundant radioactivity was not synthesized in the explosion or because the inner ejecta layers 
remained bound and fell back unto the compact remnant.  The fallback process is not well understood; it has mostly been
studied in parameterized 1-D models with an artificial 
inner boundary condition (e.g., \citet{Zhang_2008}, but see \citet{ugliano12}).  Fallback is expected to be most significant in low-energy explosions,  but it can also  be substantial in more energetic
SNe if a strong reverse shock propagates inward and decelerates the 
inner layers of ejecta \citep{Chevalier_1989, Zhang_2008, dexter13}.    If the progenitor experienced a heavy mass loss episode just prior to explosion (as we discuss below) the interaction of the SN with the CSM could produce a reverse shock which may  promote fallback of the inner ejecta.  

The second  implication of our analysis is that the progenitors of some stripped-envelope SNe may have 
surprisingly large initial radii, perhaps $R \sim 20~R_\odot$ for SN~2010X and perhaps $R \gtrsim 100 R_\odot$ for SN~2002bj. This is considerably larger the expected radii of most Wolf-Rayet (WR) stars, 
$R_0 \sim {\rm few}~{\rm R}_\odot$.   Recent stellar evolution models suggest that some stars with helium envelopes can have radii on the order
of $10~R_\odot$ \citep{yoon10}.  However, for SN~2010X we favor a composition dominated by oxygen, not helium.
In the absence of some refinement of our understanding of stellar evolution, the large inferred radius  presumably requires some mechanism to puff up an oxygen star prior to explosion. 

One compelling explanation for the large  radius is mass loss shortly before explosion.   
There are both observational and theoretical indications that instabilities can
 drive significant outflows from massive stars during the late stages of evolution \citep{woosley07, smith_2011, quataert12,Smith_Arnett_2013}. 
Most studies have focused on  mass loss episodes occurring $\sim$years prior to core collapse, which could explain
the most luminous SNe observed \citep{Smith_2007, gal-yam12, quimby12}.  
If the expelled mass expands at roughly the escape velocity of a compact star
($\sim 1000~\kms$), it will form a circumstellar shell at rather large radii, $R \sim 10^{15}$~cm. 
This shell, if it is optically thick,  sets the effective ``radius'' of the progenitor, which can produce a very bright SN light curve ($L \sim 10^{44}$~ergs, see Equation~\ref{eq:L_sn}).  

The much fainter SN~2010X could  be explained in a similar way if a circumstellar shell was located at a smaller radius, $R \sim 20~R_\odot$.  This would imply a much shorter time delay ($\sim 1$~day) between mass loss and explosion.  In fact, this dichotomy of timescales could be tied to the basic  nuclear physics of massive stars; the timescale
of the oxygen burning phase is $\sim 1$ year, while that of the silicon burning phase is $\sim 1$~day.  If mass is lost during silicon burning at $\sim 1000~\kms$, the resulting circumstellar material would have reached a radius of $10^{12}-10^{13}$~cm  at the onset of core collapse a day or so later,  setting the stage for a SN~2010X-like event.
Because the shell is relatively close to the explosion site, and any observational indications of interaction (e.g., narrow emission lines) would only be visible for a short time ($\sim 1$~hour) after explosion.

\citet{quataert12} and \citet{shiode13} \citep[see also][]{Smith_Arnett_2013} have presented an explicit mechanism for mass loss related to  core fusion.  They show that waves excited by vigorous convection during oxygen burning can (under certain circumstances)  propagate  
through the star and deposit of order $10^{47}-10^{48}$~ergs near the surface, sufficient to unbind $\sim10~\msun$ in a red supergiant.  They also show that for stripped-envelope stars, the sound-crossing time is short enough ($\lesssim 1$~day) for this mechanism to operate during silicon burning  as well.

To explain the light curve of SN~2010X, the mass in the circumstellar shell (or inflated envelope) must be substantial enough to provide a sufficiently long light curve, which may require $M_{\rm csm} \gtrsim 0.5~\msun$ (using Equation \ref{eq:t_sn} with a timescale of $\sim 20$ days, $E\sim 1~{\rm B}$, and $R\sim20~{\rm R}_\odot$).  The energy needed to drive this amount of material from a compact star is   $\gtrsim 5 \times 10^{48}$~ergs.    This is $\sim 1\%$ of the total energy released during silicon burning, but most of the fusion energy is lost to neutrinos.  In the specific models of \citet{shiode13}, wave driven mass loss in stripped envelope stars only ejects $\lesssim  0.01~\msun$ of material in the silicon burning phase.  However, more efficient mechanisms for tapping the burning energy may be possible \citep{Smith_Arnett_2013}.  The SN~2010X-like events  provide  motivation to better understand the very late phases of Type~Ib/c SN progenitors.

There may be other ways to expand the effective radius of a stripped-envelope progenitor, perhaps
related to stellar mergers or a common envelope phase in a binary system \citep[e.g.,][]{chevalier12}.
Alternatively, a large effective radius could be due to reheating of the SN remnant after it has expanded for a brief time.  \citet{dexter13}, for example, explore the possibility that the input of accretion power of a central black hole,  fed by fallback material, can produce a diversity of SN light curves.

Though they have previously been seen as weak explosions, our analysis suggests that faint, fast SNe like SN~2010X  may have more in common with the most luminous SNe in the Universe, namely the superluminous SNe powered by interaction with circumstellar material.  In both cases, the  progenitors may be  massive stars that have experienced heavy mass loss  just prior to explosion.  The main distinction in the observed light curve may simply be in the timing of the main pre-SN mass loss episode.  Further detailed modeling is needed to investigate the dynamics of the interaction and the variety of outcomes, and to determine whether realistic progenitors can produce core-collapse SNe of this type.

\begin{deluxetable}
	{ccc} 
	\tabletypesize{\scriptsize}
	 \tablecaption{\label{}} 
	 \tablewidth{0pt} 
	 \tablehead{ \colhead{species} & 
	\colhead{inner abundance\tablenotemark{a}} &
	\colhead{outer abundance\tablenotemark{b}} }
	 \startdata 
	 H & 1.3441e-14 &  ---\\
	 He & 7.2524e-13 & 9.8671e-01 \\
	 Li & --- & 1.0043e-08 \\
	 Be & --- & 1.7418e-10 \\
	 B & --- & 4.9905e-09 \\
	 C & 1.3561e-02 & 2.2179e-03 \\
	 N & 2.3119e-09 & 7.1266e-04 \\
	 O & 6.5954e-01 & 5.9003e-03 \\
	 F & --- & 3.9087e-07 \\
	 Ne & 1.5747e-01 & 1.1163e-03 \\
	 Na & 3.4553e-05 & 3.4553e-05 \\ 
	 Mg & 2.7580e-02 & 6.4116e-04 \\
	 Al & --- & 5.8630e-05 \\
	 Si & 9.5727e-02 & 7.3099e-04 \\
	 P & --- & 6.7316e-06 \\
	 S & 3.9061e-02 & 3.7317e-04 \\
	 Cl & --- & 4.7954e-06 \\
	 Ar & 4.4617e-03 & 9.6441e-05 \\
	 K & --- & 3.7829e-06 \\
	 Ca & 1.2866e-03 & 6.5889e-05 \\
	 Sc & --- & 3.9365e-08 \\
	 Ti & 2.8182e-06 & 2.8182e-06 \\
	 V & --- & 3.6517e-07 \\
	 Cr & 2.9462e-06 & 1.6984e-05 \\
	 Mn & --- & 1.4106e-05 \\
	 Fe & 1.3033e-03 & 1.2132e-03 \\
	 Co & 3.4272e-06 & 3.4272e-06 \\
	 Ni & 7.0975e-05 & 7.0975e-05 \\
	\tableline 
\enddata 

\tablecomments{Mass fractions used for the composition in the radiative transport models. The boundary between inner and outer zones is at $10^9$ cm. For some isotopes in the inner layer, the abundance was increased to solar. \label{t:comp}} 
\tablenotetext{a}{ \citet{woosley02} oxygen/neon-rich composition.} 
\tablenotetext{b}{ Solar composition from \citet{Lodders_2003} with all hydrogen converted to helium.}
\end{deluxetable}

\clearpage

\section*{Acknowledgments}

We would like to thank Maria Drout, Alex Heger, Mansi Kasliwal, Ehud Nakar, Christian Ott, Tony Piro,
 Dovi Poznanksi, Josh Shiode, Alicia Soderberg, and Eliot Quataert
for useful discussion.
This work was supported by an NSF Astronomy and Astrophysics Grant (AST-1109896) and by an NSF Division of Astronomical Sciences collaborative research grant AST-1206097.
I.K is supported by the Department of Energy National Nuclear Security Administration Stewardship Science Graduate Fellowship.
D.K is supported in part by a Department of Energy Office of Nuclear
Physics Early Career Award, and by the Director, Office of Energy
Research, Office of High Energy and Nuclear Physics, Divisions of
Nuclear Physics, of the U.S. Department of Energy under Contract No.
DE-AC02-05CH11231.   We are grateful for computing time made available
the National Energy Research Scientific Computing Center, which is supported by the Office of Science of the U.S. Department of Energy under Contract No. DE-AC02-05CH11231.

\bibliographystyle{mn2e}

\bibliography{sn10x_2}

\end{document}